\documentclass{article}

\usepackage{microtype}
\usepackage{graphicx}
\usepackage{subfigure}
\usepackage{booktabs}
\usepackage{multirow}
\usepackage{hyperref}
\usepackage{xurl}
\hypersetup{breaklinks=true}

\usepackage[accepted]{icml2024}

\usepackage{amsmath}
\usepackage{amssymb}
\usepackage{mathtools}
\usepackage{amsthm}
\usepackage[capitalize,noabbrev]{cleveref}

\theoremstyle{plain}

\theoremstyle{definition}

\theoremstyle{remark}

\icmltitlerunning{End to End Learning StatArb}

\begin{document}

\twocolumn[
\icmltitle{End-to-End Policy Learning of a Statistical Arbitrage Autoencoder Architecture}

\icmlsetsymbol{equal}{*}

\begin{icmlauthorlist}
\icmlauthor{Fabian Krause}{yyy}
\icmlauthor{Jan-Peter Calliess}{yyy}

\end{icmlauthorlist}

\icmlaffiliation{yyy}{Department of Engineering, University of Oxford, Oxford, United Kingdom}

\icmlcorrespondingauthor{Fabian Krause}{lina3477@ox.ac.uk}

\icmlkeywords{Machine Learning, Autoencoder, Statistical Arbitrage, Mean Reversion, Finance, ICML}

\vskip 0.3in
]

\printAffiliationsAndNotice{}

\begin{abstract}
In Statistical Arbitrage (StatArb), classical mean reversion trading strategies typically hinge on asset-pricing or PCA based models to identify the mean of a synthetic asset. Once such a (linear) model is identified, a separate mean reversion strategy is then devised to generate a trading signal. With a view of generalising such an approach and turning it truly data-driven, we study the utility of Autoencoder architectures in StatArb. As a first approach, we employ a standard Autoencoder trained on US stock returns to derive trading strategies based on the Ornstein-Uhlenbeck (OU) process. To further enhance this model, we take a policy-learning approach and embed the Autoencoder network into a neural network representation of a space of portfolio trading policies. This integration outputs portfolio allocations directly and is end-to-end trainable by backpropagation of the risk-adjusted returns of the neural policy. 
Our findings demonstrate that this innovative end-to-end policy learning approach not only simplifies the strategy development process, but also yields superior gross returns over its competitors illustrating the potential of end-to-end training over classical two-stage approaches.

\end{abstract}

\section{Introduction}
\label{introduction}
Quantifying relationships between financial assets using statistical techniques has long captivated both researchers and practitioners. The concept of exploiting mean reverting prices in pairs trading, where one asset is traded against another upon divergence from their established relationship, laid the groundwork for this field. Significant excess returns identified in pairs trading strategies between 1962 and 2002, as reported by \citet{PTPerfRelValue}, highlight the potential of such approaches. A systematic review of pairs trading research is provided by \citet{PairsReview}. Our focus extends beyond pairs to the co-movement of groups of stocks or their underlying factors. \citet{StatArbUS} formally generalised the idea of pairs to StatArb trading that is said to have been developed in the mid 1980's in Morgan Stanleys trading group around Nunzio Tartaglia \citep{PTPerfRelValue}.

StatArb trading assumes an asset pricing model incorporating various statistical or fundamental factors to describe an asset's price movement. There is a multitude of asset pricing options to choose from depending on context and modelling goals. The idea of StatArb is that deviations of an asset's return from the model's explained return are temporary and will revert to the model returns. This necessitates modelling the reversion time series process and combining various asset signals into a tradable portfolio. 

Principal Component Analysis (PCA) is the work horse for uncovering the statistical factors in asset pricing models. Our first contribution is replacing PCA with an Autoencoder to derive tradeable residuals. It is well known since \citet{Oja} that Autoencoders can replicate and extend PCA's capabilities by including non-linearities. We compare the performance of trading strategies based on our simple Autoencoder asset pricing models against the literature benchmark models based on PCA as well as based on the classic fundamental asset pricing models using Fama French factors \citep{ChoosingFactors}. This performance comparison of Autoencoder based asset pricing models against PCA and Fama French has, to the best of our knowledge, not been systematically explored in the literature.

The multitude of modelling decisions in the StatArb process introduces substantial modelling risk. Asset pricing models are constructed to explain asset returns, not to find residuals that exhibit mean reversion characteristics. The mean reversion characteristic is a by-product that is exploited by traders. Recognising this, we propose embedding an Autoencoder in a neural network representing a trading policy, trained end-to-end with a loss function that optimises both portfolio return representation and risk-adjusted returns. We study the performance of different variants of the architecture and competing methods on historical US equity returns, demonstrating the advantages of an end-to-end training approach over traditional methods.

Our methodology not only learns a statistical factor model from data but also extracts residuals within the network, leading to a portfolio that is optimised for StatArb trading through a risk-adjusted performance policy. This architecture, as we will demonstrate, surpasses prevalent benchmark solutions in StatArb trading, offering improvements in terms of return before cost and reducing modelling risk at every stage of the process. 

\section{Related Work}
The concept of StatArb trading, as we define it, finds its roots in the foundational work of \citet{StatArbUS}. Their basic idea is that price changes follow the differential equation
\[\frac{dP_t}{P_t} = \alpha dt + \sum_{j=1}^n \beta_j F_t^{(j)} + d X_t \]
where $X_t$ is a stationary mean reverting process, referred to as the cointegration residual, $P_t$ is the price of a security at time t, $\alpha$ a drift term, and $\beta_j$ factor exposures for the risk-factors $F_t^{(j)}$. Statistical arbitrage is the generalisation of pairs-trading in which $n=1$. They show how modelling the risk factors as industry ETFs or principal components and modelling the residual process as a mean reverting Ornstein-Uhlenbeck (OU) process, one can derive a profitable trading strategy.  \citet{StatarbRiskControl} further refine the idea to include risk controls depending on the speed of mean reversion of residuals.

There is a proliferation of factors and asset pricing models in the financial literature \citep{HarveyExpected,TamingZoo}. The seminal contributions in this space were performed by \citet{CAPMSharpe} developing the Capital Asset Pricing model, \citet{APT} introducing multi-factor models in his arbitrage pricing theory, and \citet{CommonFrench} defining the Fama French factors.
Lately, the state of the art in factor models was pushed by combining exogeneous and latent factors using an Instrumented Principal Component Analysis (IPCA) proposed by \citet{IPCA}. \citet{AutoencoderIPCA} built upon that work and introduced a novel autoencoder architecture that incorporates IPCA at its foundation. While similar in technique to our work, their work focuses on explaining the cross-section of returns rather than constructing a trading strategy that exploits model deviations. 

Further, \citet{DeepLearningStatarb} explored the use of a Transformer-based model for StatArb portfolio construction, processing residual series from various asset pricing models. While this model integrates residual extraction and portfolio construction, it leaves the asset pricing model selection at the discretion of the user. 

Notably, \citet{TwoML} also proposed using Autoencoder-generated residuals for StatArb strategies, employing convolutional Autoencoders. However, their approach, unlike ours, does not integrate all stages of the StatArb process and lacks a comprehensive evaluation against a range of data and competing models.

Our work seeks to contribute to this evolving field by presenting an integrated approach that leverages Autoencoders for residual generation, embedding these within a comprehensive trading strategy that is optimised end-to-end. This method aims to address gaps in current methodologies, offering a holistic solution to StatArb trading that encompasses model generation, risk management, and strategy optimisation.

\section{Statistical Arbitrage}
The following sections lay out the building blocks for a successful StatArb trading strategy and the assumptions that such a strategy is based upon.
\subsection{Asset Pricing Models}
Asset pricing models rest on the assumption of no arbitrage - the premise that all excess returns are attributable to risk factors. This assumption implies that any unexplained excess return signals an overlooked factor in the model. Historically, stock price anomalies were incrementally recognised and integrated into these models, a process formalised using a stochastic discount factor by \citet{IPCA}. The general formulation of an asset pricing model is:
\[r_{i,t+1} = \alpha_{i,t} + \beta_{i,t}^T f_{t+1} + \epsilon_{i, t+1}
\]
for $\beta_{i, t} \in \mathbb{R}^{k x 1}$, the factor exposures of stock $i$ to the selected $k$ factors that exhibit returns $f_{t+1} \in \mathbb{R}^{k x 1}$ at time $t$ and a residual $\epsilon_{i, t+1}$ that is not explained by the given factor returns and therefore is stock specific. Further, $E_t[\epsilon_{i,t+1}] = E_t[\epsilon_{i,t+1} f_{t+1}] = \alpha_{i,t} = 0$ if the stochastic discount factor accounts for all statistically significant anomalies. Model returns are given by $\beta_{i, t}^T f_{t+1}$. They are what, according to the model, a stock should have returned at a given time with given factor returns for that time. 

There are primarily two methods to model the factor structure in asset pricing models:
\begin{itemize}
\item Observable Exogeneous Factors: Either the factor loadings are predefined, e.g. as firm characteristics, or the factor returns are given, with betas and alphas estimated via cross-sectional regressions.
\item Latent Factors: Assuming the factors to be latent and estimating them via PCA or similar latent factor analysis techniques.
\end{itemize}

In the former case when factor returns are given, loadings $\beta_{i, t}$ for stock $i$ at each time step $t+1$ are estimated using a cross-sectional regression with a lookback window of $T$. Residuals for time step $t+1$ are retrieved via:
\[
 \epsilon_{i,t+1} = r_{i,t+1} - \beta_{i, t}^T f_{t+1}
\]
In matrix notation, the regression for each time step $t+1$ can be denoted as
\[
R_{t-T:t} = F_{t-T:t} \beta_{t}^T + \epsilon_{t-T:t}
\]
with stock returns $R \in \mathbb{R}^{T x N}$ for N stocks, factor returns $F \in \mathbb{R}^{T x k}$, factor loadings $\beta \in \mathbb{R}^{N x k}$, and residuals $\epsilon \in \mathbb{R}^{T x N}$. The out of sample residuals for the model are retrieved via
\[
\epsilon_{t+1} = R_{t+1} - \beta_{t}^T  F_{t+1}
\]
with $\epsilon_{t+1} \in \mathbb{R}^{1 x N}$, $R_{t+1} \in \mathbb{R}^{1 x N}$, $\beta_t \in \mathbb{R}^{k x 1}$, $F_{t+1} \in \mathbb{R}^{k x 1}$.

The case of using a latent factor structure is similar to the benchmark PCA methodology explained in section \ref{PCA}.

\subsection{Arbitrage Signal Extraction}
In StatArb trading, a core assumption is that the expected value of the residuals is zero, $E_t[\epsilon_{i,t+1}] = 0$. The belief is that residuals are mean reverting. However, the difficulty lies in accurately timing the mean reversion for a profitable trading strategy. Therefore, one needs to employ time series prediction techniques on the univariate residual process to generate signals to act upon. 

We denote a generic time series extraction technique as $\phi$, the signal at the end of time $t$ is $\phi(\epsilon_{t-\tau:t}) \in \mathbb{R}^N$ for a lookback window of residuals $\tau$ and N stocks in the universe at time $t$.

This extraction allows to make informed predictions about the likely future behaviour of stock prices based on their past residual patterns. By doing so, the aim is to better time trading strategies, exploiting the predicted mean reversion of these residuals.

\subsection{Arbitrage Portfolio Construction}
The final stage of the StatArb process uses the time series predictions $\phi(\epsilon_{i,t-\tau:t})$ at time $t$ to construct a portfolio of weights, represented as $w_t$. These can be optimised using classic portfolio optimisation techniques including constraints on holdings, factor exposures, and targeting risk levels. The simplest combination, and the one we use throughout to highlight and compare the modelling aspects of a statistical arbitrage trading pipeline, is to target constant leverage in the portfolio, i.e. 
\[
w_{i, t} = \frac{\phi_i(\epsilon_{i, t-\tau:t})}{|| \phi_i(\epsilon_{i, t-\tau:t} ||_1}.
\]
This ensures a target leverage of 1, with the syntax $|| . ||_1$ denoting the $L_1$ norm.

Weights generated in this way require stock return information up to the end of time $t$. We assume we can take positions using this weight at the beginning of time $t+1$ and capture the return from the beginning to the end of $t+1$. This assumption is not unrealistic given the liquidity of the considered universe of stocks and the state of electronic markets.

The return of a portfolio $p$ at time $t+1$ given weights $w_{i,t}$ is
\[
r_{p, t} = \sum_{i=1}^N w_{i, t} r_{i, t+1}.
\]

\section{Methodology}
\subsection{Data}

CRSP\footnote{\url{https://www.crsp.org}} stock data was used in the experiments from January 2000 to December 2022. While we recognise that the asset pricing literature uses longer lookbacks and historic data before 2000, research results around portfolio returns on earlier data are seldom reproducible in recent years. In particular, much of the statistical arbitrage literature generates great strategy returns up until 2007 but fails to generalise to similar results post 2007. Furthermore, the aim of asset pricing research is to explain stock returns whereas we present a trading strategy. 

We select only US traded companies and restrict our analysis to the primary listing of a given company. Prices are adjusted for dividends and stock splits to run experiments on an adjusted total return series. This leaves a total of 13.610 companies in our data set and 5.787 daily observations.

Furthermore, for our benchmark analysis, we use Fama French factor returns for the same time period\footnote{Data can be found at \url{http://mba.tuck.dartmouth.edu/pages/faculty/ken.french/data_library.html\#Research}}. The data contains the daily risk-free rate, returns for the market factor, size factor, value factor, profitability factor, investment factor, and momentum factor.

In line with the literature on financial asset pricing, all our performance analyses are done on excess returns, i.e. returns in excess of the risk-free rate.

\subsubsection{Universe of Eligible Stocks}
Statistical arbitrage trading tends to incur high transaction costs due to a high volume of daily turnover in a large universe of stocks. We restrict our universe by removing any stocks with a close price lower than \$5, a 20-day median rolling market capitalisation of below \$1bn, or a 20-day median rolling trading volume below \$1m. We update the filter of stocks to include at the end of each calendar month. 

This universe construction restricts the number of stocks included over the whole lookback period from 13.611 companies to 5.188. On average, 1.470 stocks are included each month with the number of included stocks increasing in recent years. 

\subsection{Benchmarks for Asset Pricing Models}
\subsubsection{Fama French}
To extract residuals from the Fama French model, we employ a daily regression process. This involves regressing the stock returns against the returns of Fama French factors using the past $T_{FF}=60$ days. The factor loadings for our stock universe are projected to the next day and model returns are derived from which the residuals are retrieved. Our analysis encompasses the following factor models, each selected for its relevance in capturing different market dynamics:
\begin{itemize}
\item CAPM: 1 factor representing the market, commonly termed the capital asset pricing model (CAPM).
\item FF 3: Fama French 3 factor model with market, size and value factors.
\item FF 5: Fama French 5 factor model including profitability and investment factors.
\item FF 5 Mom: Fama French 5 factor model plus a momentum factor.
\end{itemize}

\subsubsection{Principal Component Analysis} \label{PCA}
Residuals from the PCA approach are retrieved according to the methodology of \citet{StatArbUS}. We use factor models keeping 1, 3, 5, 8, 10, 15, and 20 principal components in line with \citet{DeepLearningStatarb}. 

Each day we use standardised returns of the past $T_{\tau}$ = 252 trading days, i.e. one trading year, to estimate the empirical correlation matrix. For $\tau = 0, ..., T_{\tau}-1$ the standardised returns $Z_{i, t-\tau}$ are
\[
Z_{i, t-\tau} = \frac{r_{i, t-\tau} - \mu_{i, t-T_{\tau}:t}}{\sigma_{i, t-T_{\tau}:t}}
\]
where $\mu_{i, t-T_{\tau}:t}$ denotes the sample mean of returns for the period of $t-T_\tau$ to $t$ and $\sigma_{i, t-T_{\tau}:t}$ the sample standard deviation of returns for the same period. These standardised returns form the empirical correlation matrix
\[
C_{t} = \frac{1}{T_{\tau}-1} Z_t^T Z
\]
with $Z_t \in \mathbb{R}^{T_{\tau} x N}$ the returns preceding and including time $t$ for all N stocks. $C_t$ is used to extract eigenvalues and eigenvectors.

Keeping only a subset of the principal components as factors, we calculate returns of the \textit{eigenportfolios} \cite{StatArbUS} as 
\[
F_{t, j} = \sum_{i=1}^N \frac{v_i^{(j)}}{\sigma_i} r_{i, t}
\]
with $v_i^{(j)}$ the $j$-th principal component's entry for stock $i$ and $\sigma_i$ stock $i$'s volatility. These are the PCA factor returns.

Once factor returns are calculated for each time step $t$, we run a regression akin to the Fama French case to estimate loadings of stocks on the latent factors using the past $T_{FF}=60$ days. This regression uses non-standardised returns and yields $\beta_t$ estimates to calculate residuals for time $t+1$. Note that time step $t+1$ was not used during the PCA or regression steps. Similar to the cited research, we also only use stocks with no missing observations during the lookback window. 

\subsection{Autoencoder Residual Generation}
Our first contribution entails the extension of the PCA setup for creating residuals by using Autoencoders in a StatArb framework. All of our tested Autoencoders use an output layer that contains a hyperbolic tangent activation and a MSE loss function optimised with an Adam optimiser for 10 epochs to find weights that best model the standardised input returns. We use the past $T_\tau$ trading days to train the Autoencoder, mirroring the PCA model setup, using only stocks with no missing values for standardised returns throughout the lookback period. Standardised returns are capped at 3 standard deviations to prevent outliers to distort the model's training.

Our architecture can be summarised by the following equations and loss function $L$. Let $Z$ denote the standardised returns as defined earlier, then:

\begin{align*}
F &= enc(Z) = f_e(W^{(0)} Z + b^{(0)}) \\
\hat{Z} &= dec(F) = f_d(W^{(1)} F + b^{(1)}) \\
L &= MSE(Z, \hat{Z}) = \frac{1}{n} \sum_{i=1}^n (Z_i - \hat{Z}_i)^2
\end{align*}
where $f_e$ is the encoder activation function, the biases $b$ are optional depending on which variant is run, $f_d$ is the decoder activation function of the network, in our case fixed to a hyperbolic tangent function, and MSE is the mean squared error loss function. 

A new model is fitted every day. The variation is in the number of layers, activation of the encoder part of the network, whether to use biases in the linear layer generation, and whether to use dropout layers. Additionally, we also test several distinct options of extracting residuals from the trained network: 
\begin{enumerate}
\item Subtracting the Autoencoder predicted standardised returns from the actual standardised returns, i.e.\\
\[ \epsilon_{i, t} = Z_{i, t} - \hat{Z}_{i, t}.
\]
\item Extracting \textit{factor returns} for the past 60 days by predicting the latent layer in the network using just the encoder part of the Autoencoder and running a regression of factor returns against actual returns to extract factor loadings. We then use the final day's standardised return (that was not included in running the regression) to get an encoder prediction of factor returns. These factor returns are multiplied by the estimated loadings and subtracted from the actual returns to retrieve the residuals. \\

The regression to estimate the factor loadings $\beta$ can be summarised as:

\[
R_{t-60:t-1} = \beta \ enc(Z_{t-60:t-1}) + \epsilon_{t-60:t-1} \]
which leads to residuals using the estimated $\beta$:
\[
\epsilon_{t, i} = r_{i, t} - \beta \ enc(Z_t).
\]

\item Using the procedure as in the second option but instead of using standardised returns to retrieve factor return estimates, we use volatility scaled returns to mirror the setup of \textit{eigenportfolios} in \citet{StatArbUS}. The scaling of returns uses the backward-looking standard deviation of the past $T_\tau=252$ trading days. \\
The regression changes to 
\begin{align*}
R_{t-60:t-1} &= \beta \ enc(S_{t-60:t-1}) + \epsilon_{t-60:t-1}\\
\epsilon_{t, i} &= r_{i, t} - \beta \ enc(S_t)
\end{align*}
with $S_{i, t} = \frac{r_{i, t}}{\sigma_{i, t-T_{\tau}-1:t-1}}$.

\end{enumerate}
The final two options and especially the volatility scaled returns option are inspired by the relationship between Autoencoders and PCA as well as the use of PCA in StatArb trading. 

Our choice for using the hyperbolic tangent function in the decoder stems from the fact that we are trying to recreate standardised stock returns in the decoder. For this comparison we keep the number of hidden nodes fixed at 20 for all options and variants.

The variants that are run for each residual option are labelled and defined in table \ref{ae-variants}.

\begin{table}[t]
\caption{Autoencoder architecture choices for residual generation. \textit{TANH} denotes the hyperbolic tangent activation function and \textit{RELU} the rectified linear unit.}
\label{ae-variants}
\vskip 0.15in
\begin{center}
\begin{small}
\begin{sc}
\begin{tabular}{lcccr}
\toprule
Variant & Activation & Bias & Dropout &  Layers \\
\midrule
 0 & tanh & True & 0.25 & 1 \\ 
 1 & tanh & True & 0.25 & 3 \\
 2& tanh & True & False & 1 \\
 3 & tanh & False & False & 1 \\
 4 & tanh & False & False & 3 \\
 5 & relu & True & 0.25 & 1 \\
 6 & relu & True & 0.25 & 3 \\
 7 & relu & True & False & 1 \\
 8 & relu & False & False & 1 \\
 9 & relu & False & False & 3 \\
\bottomrule
\end{tabular}
\end{sc}
\end{small}
\end{center}
\vskip -0.1in
\end{table}

\subsection{Benchmarks for Arbitrage Signal Extraction}
Once residuals are calculated using the discussed asset pricing models, time series techniques are applied on the residuals' processes to extract signals that form the inputs to the portfolio construction step. These time series signal extraction techniques are applied univariately, i.e. for each stock's residuals separately.

\subsubsection{Ornstein Uhlenbeck}
The most prevailing approach in the literature for modelling the residual time series is using a mean reverting Ornstein-Uhlenbeck (OU) process. We follow the methodology of \citet{StatArbUS} and estimate its parameters assuming a discretised version to model the idiosyncratic residual $dX_i(t)$ within the stochastic differential equation
\[
dX_i(t) = k_i (m_i - X_i(t)) dt + \sigma_i d W_i (t), \quad k_i > 0.
\]
Like in \citet{StatArbUS} \textit{s-scores} are calculated as
\[s_i = \frac{X_i(t) - m_i}{\sigma_{eq,i}}
\]
with $\sigma_{eq,i} = \frac{\sigma_i}{\sqrt{2 k_i}}$. Buy and sell decisions are driven by these s-scores and using historically chosen thresholds for opening and closing positions. We follow the already cited work in this area and use a threshold of -1.25 (1.25) to open a long (short) position and -0.5 (0.75) to close the open long (short) position. Moreover, we only take positions in stocks that show a high $R^2$ value, representing confidence in the regressions, and chose a threshold of 0.25. This technique is described in \citet{StatarbRiskControl}.

\subsubsection{Artifical Neural Network}
Additionally, we follow the example of \citet{DeepLearningStatarb} and use a 3-layered feed forward network with 5 input parameters to extract signals from the OU estimation output statistics. These parameters are part of the discretisation process described in the appendix of \citet{StatArbUS} and represent a summary of the residual process at each time step $t$.

The feed forward networks make univariate predictions using all 5 parameters as inputs, transforming them via 3 layers of each 5 nodes with a $relu$ activation function and a dropout layer with a dropout weight of 0.25. The output is a single value from a linear layer. We train the network using a Sharpe ratio policy on batches of 1.000 outputs using the Adam optimiser and 5 epochs of training.

Our backtesting methodology is the same as the one described by \citet{DeepLearningStatarb}:
\begin{itemize}
\item At each time t, we extract OU parameters from the residual time series using a 60 day lookback regression.
\item To train the model, 1.000 days of history of OU parameters are used for all stocks and optimised using a Sharpe ratio objective.
\item The next 125 days are used out of sample to make predictions and create portfolios for performance evaluation before a new model is trained.
\end{itemize}

Training start dates are adjusted across models to accommodate the varying length of lookbacks required. We refer to this extraction technique as \textit{OU+FFN} in the results section.

\section{Autoencoder Statarb Policy} \label{AEPol}
In the realm of statistical arbitrage trading, as delineated in the preceding sections, the efficacy heavily relies on employing an asset pricing model that accurately captures the variability in stock returns. The plethora of asset pricing models and factors makes the choice for one to be used at best a data mining exercise that is prone to overfit. Moreover, the modeler is tasked with choosing an apt signal extraction technique and adjusting portfolio weights accordingly.

To address these challenges, we propose an integrated Autoencoder training procedure that facilitates learning a factor model that is tailored for StatArb trading. This way, we combine the asset pricing model estimation with the estimation of StatArb parameters into a unified, cohesive model. This integration allows end-to-end decision making and circumvents the need for external asset pricing models and time series prediction techniques. 

The encoder consists of one layer with the number of $l$ nodes as specified in the results section. We use a similar number of hidden units as in the PCA model to simplify comparison of the methods. It uses a rectified linear unit as the activation in its hidden layer: 

\[F = enc(Z) = relu(W^{(0)} Z + b^{(0)}) \]

where $Z\in \mathbb{R}^{N x 1}$, $W^{(0)} \in \mathbb{R}^{l x N}$, and $b^{(0)} \in \mathbb{R}^{l x 1}$.
The decoder also only consists of one layer, denoted as $dec$,  with the same dimension as the input data, i.e. the number of included stocks $N$. It uses a hyperbolic tangent activation function to allow the network to best represent the input data:

\begin{align*}
\hat{Z} &= dec(F) \\
&= tanh(W^{(1)}F + b^{(1)} ) \\
&= tanh(W^{(1)} (relu(W^{(0)} X + b^{(0)})) + b^{(1)} )
\end{align*}
where $F\in \mathbb{R}^{l x 1}$, $W^{(1)} \in \mathbb{R}^{N x l}$, and $b^{(1)} \in \mathbb{R}^{N x 1}$.

The decoded values are used in the loss function to minimise the MSE between inputs and decoded outputs to allow the network to learn a useful latent space representation without loss of expression.

To allow the architecture to be geared towards StatArb trading, we add two additional layers to the network that feed into the trading policy. The first layer calculates the difference between decoded returns and the original input standardised returns. This yields residuals akin to the ones generated by traditional factor models. The second layer takes these residuals as inputs and builds a portfolio of weights, $w_t$, by using a hyperbolic tangent activation function with no bias term:

\[w_t = tanh(W^{(2)} (\hat{Z}_t - Z_t)), \quad W^{(2)} \in \mathbb{R}^{N x N}. \]

Its reuse of the inputs $Z$ makes this layer similar to residual or skip connections in classic deep learning literature.

The optimisation problem for this network includes minimisation of a Sharpe ratio policy. We also ensure that the sum of the final absolute positions equals 1 to prevent accidentally leveraging the portfolio, i.e. $w_t = \frac{w_t}{|| w_t ||_1}$. By introducing $\lambda$ as a gearing parameter, to weight the MSE and the Sharpe ratio components of the loss function differently, the objective is formalised as follows:

\begin{equation}
\begin{aligned}
\min_{\theta} \quad & \lambda \: MSE(Z_t, \hat{Z}_t) + (1 - \lambda) \ Sharpe(w_t, R_{t+1})  \\
\textrm{s.t.} \quad & \sum_{i=1}^n w_{i, t} = 1 \quad \forall t \\
and \quad & \theta = \{ W^{(0)}, W^{(1)}, W^{(2)},  b^{(0)},  b^{(1)}\}
\end{aligned}
\end{equation}

with network parameters $\theta$.

For the results in this paper, we keep $\lambda=0.5$. The Sharpe ratio policy is defined as:

\begin{equation}
\begin{aligned}
Sharpe &= \sqrt{252} \: \frac{\mu_{p, t}}{\sigma_{p, t}} \\
\mu_{p, t} &= \frac{1}{T} \sum_{\tau=1}^T r_{p, t-\tau} \\
\sigma_{p, t} &= \sqrt{\frac{1}{T} \sum_{\tau=1}^T (r_{p, t-\tau} - \mu_{p, t})^2 }
\end{aligned}
\end{equation}
with $r_{p, t} = \sum_{i=1}^n w_{i, t} \: r_{i, t+1}$.
The factor $\sqrt{252}$ annualises the Sharpe ratio as we use daily estimates of mean and standard deviation of returns. This policy could further be extended to incorporate portfolio constraints.

\subsection{Training}
In line with the benchmark models, we train a new model every trading day using the past $T_\tau=252$ days of standardised returns, capped at 3 standard deviations, as an input to the Autoencoder model. The Sharpe ratio objective is evaluated on actual returns of the next day. For each model, we only use past data as training inputs and project the model forward to retrieve the model's performance on the next day out of sample. 

Models are trained for 10 epochs using the Adam optimiser and a 0.001 learning rate. 

\section{Results} \label{results}

Our first key contribution is the systematic comparison of our basic Autoencoder models with baseline asset pricing models (PCA and Fama French) for residual generation in the StatArb process. Table \ref{tab-ou-comp-high} shows results for the best performing models between our basic Autoencoder variants (as defined in table \ref{ae-variants}) and the other asset pricing models using OU signal extraction. A full comparison of all models can be found in table \ref{tab-ou-comp} in the appendix. It is noteworthy that while the Autoencoder models do not surpass PCA in terms of Sharpe ratio, the regression-based models deliver competitive, and in some cases superior, returns when compared to the best-performing PCA models, albeit with increased risk, primarily due to higher turnover and daily architectural changes. Future iterations could potentially mitigate this by introducing penalties for frequent model modifications.

Results using OU+FFN extractions, which generally underperformed, are detailed in table \ref{tab-ou-ffn-app} in the appendix. Our analysis here was limited to the second methodology of residual extraction through cross-sectional regression, as it emerged as the most effective.

\begin{table}[t]
\caption{Performance comparison of asset pricing models against our simple Autoencoder models using Ornstein-Uhlenbeck Extraction. The model column describes the type of asset pricing model and signal extraction method. The variant is the sub-variant within the combination, SR is the Sharpe ratio before costs, and $\mu$ and $\sigma$ the sample mean and standard deviation of returns respectively. The numbers in the variant name of the PCA models denote the number of latent factors to keep.}
\label{tab-ou-comp-high}
\vskip 0.15in
\begin{center}
\begin{small}
\begin{sc}
\scalebox{0.9}{
\begin{tabular}{l|c|ccc|ccc|r}
\toprule
Model & Variant & SR & $\mu$ & $\sigma$ \\
\midrule
\midrule 
\multirow{4}{*}{FF OU} & CAPM & 0.13 & 1.31\% & 10.17\% \\ 
 & FF 3 & 0.3 & 2.07\% & 6.89\% \\ 
& FF 5 & 0.33 & 2.22\% & 6.78\% \\ 
 & FF 5 Mom & 0.52 & 3.39\% & 6.56\% \\ 
\midrule
\midrule
\multirow{4}{*}{PCA OU} & PCA 8 & 0.96 & 4.55\% & 4.74\% \\ 
 & PCA 10 & 0.92 & 4.3\% & 4.66\%\\ 
& PCA 15 & 0.87 & 3.94\% & 4.53\%  \\ 
& PCA 20 & 0.88 & 3.91\% & 4.43\% \\ 
\midrule
\midrule 
\multirow{4}{*}{AE OU 2} & Variant 0 & 0.14 & 2.36\% & 16.75\%  \\ 
& Variant 1 & 0.34 & 5.76\% & 16.96\% \\ 
 & Variant 2 & 0.29 & 4.84\% & 16.77\% \\ 
 & Variant 3 & 0.31 & 5.17\% & 16.75\% \\ 
\midrule
\midrule 
\multirow{5}{*}{AE OU 3}  & Variant 1 & 0.24 & 4.06\% & 17.2\%  \\ 
& Variant 2 & 0.27 & 4.44\% & 16.55\%  \\ 
& Variant 6 & 0.42 & 6.93\% & 16.52\%  \\ 
& Variant 7 & 0.32 & 5.28\% & 16.31\% \\ 
& Variant 8 & 0.32 & 5.18\% & 16.33\%  \\ 
\hline
\bottomrule
\end{tabular}}
\end{sc}
\end{small}
\end{center}
\vskip -0.1in
\end{table}

\begin{figure*}[ht]
\vskip 0.2in
\begin{center}
\centerline{\includegraphics[scale=0.45]{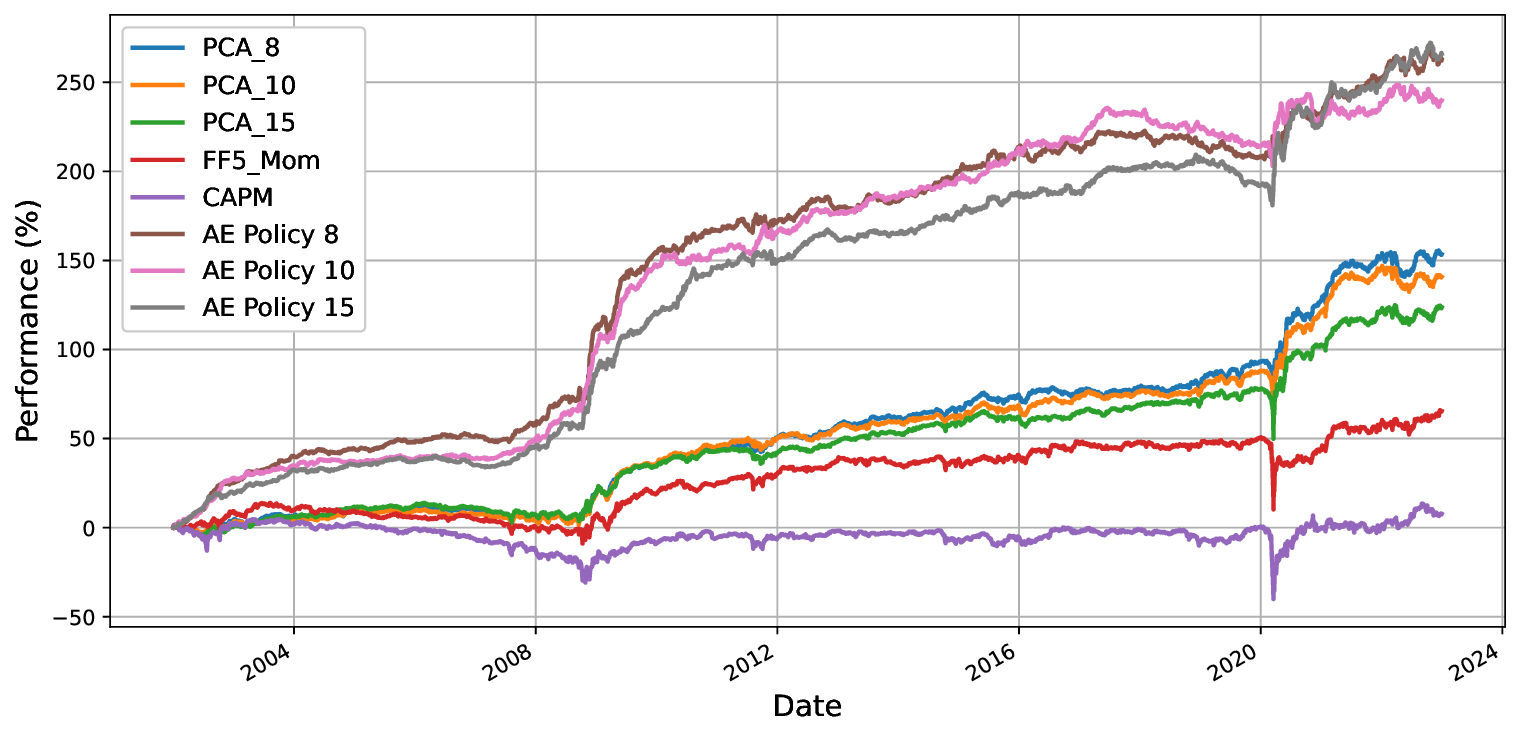}}
\caption{Performance comparison in compounded percentage returns from 2002 to end of 2022.}
\label{icml-perfcmp}
\end{center}
\vskip -0.2in
\end{figure*}

Our second major contribution is the proposal of a novel Autoencoder policy strategy, as detailed in section \ref{AEPol}. Its pre-cost performance is presented in table \ref{ae-ou-perf}, showcasing an outperformance over all benchmarks in risk-adjusted terms, with high mean returns at a risk level comparable to PCA model variants. Figure \ref{icml-perfcmp} visually contrasts these performances since 2002 focusing on the best performing variants and a diverse set of benchmark models.

The architecture's ability to learn a representation conducive to StatArb trading is notable. Interestingly, the number of latent factors in the best-performing policy models aligns with the literature's recommended range of 10 to 15, suggesting that the Autoencoder effectively learns a representation akin to qualitatively derived factors. The flexible policy learning approach also facilitates the integration of portfolio construction constraints, crucial for strategies with potential high turnover and transaction costs. Thereby, one can slow down the signal or penalise for frequent model changes. While initial analysis indicates the need for further refinement to enhance after-cost competitiveness, the model's promising gross performance underscores its theoretical potential. In a broader quantitative trading portfolio, this model could serve as an additional signal to augment overall performance.

\begin{table}[t]
\caption{Performance of our Autoencoder Policy. The dimension column defines how many latent dimensions were included in the architecture.}
\label{ae-ou-perf}
\vskip 0.15in
\begin{center}
\begin{small}
\begin{sc}
\scalebox{0.9}{
\begin{tabular}{l|c|ccc|ccc|r}
\toprule
Model & Dimension & SR & $\mu$ & $\sigma$ \\
\midrule
\midrule 

\multirow{9}{*}{AE Policy }& 3 & 1.51 & 5.56\% & 3.69\% \\ 
 & 5 & 1.42 & 5.08\% & 3.58\% \\  
 & 6 & 1.67 & 5.66\% & 3.4\% \\ 
 & 8 & 1.73 & 6.2\% & 3.58\% \\ 
 & 10 & 1.75 & 5.89\% & 3.37\% \\ 
 & 15 & 1.81 & 6.24\% & 3.46\% \\ 
 & 20 & 1.5 & 5.34\% & 3.55\% \\ 
 & 30 & 1.43 & 5.28\% & 3.68\% \\ 
 & 50 & 1.58 & 5.68\% & 3.6\% \\ 

\hline
\bottomrule
\end{tabular}}
\end{sc}
\end{small}
\end{center}
\vskip -0.1in
\end{table}

\section{Conclusion}
To the best of our knowledge, this is the first time Autoencoder architectures were systematically evaluated against the prevailing approaches to construct factor models in the context of StatArb trading. Our findings reveal that even basic Autoencoder implementations can rival the performance of established benchmark models like PCA and Fama French in a StatArb context. In the future, we aim to broaden our comparison by including IPCA approaches and the Autoencoder IPCA model introduced by \citet{AutoencoderIPCA} as additional benchmarks. Furthermore, the basic implementations that we used can be substantially improved upon by including convolutional features, using Variational Autoencoders, and additional layers. 

Our end-to-end Autoencoder architecture of learning the entire pipeline in StatArb modelling shows promising results on a pre-cost basis. It highlights the theoretically achievable performance and establishes a baseline for future work. Notably, our findings align with existing literature regarding the optimal number of latent factors in asset pricing models for StatArb trading, hinting that there is an economic or statistical foundation for this number and also bolstering confidence in our approach. Our AE policy strategies perform best when using the same number of latent factors as the optimal choice in the literature.

There is variety of future avenues to make the architecture competitive in a realistic after-cost simulation, but we already outlined initial steps for improvement in section \ref{results}. Firstly, reducing model turnover is a priority, achievable through less frequent retraining, incorporating transaction cost penalties, or modifying the Sharpe policy in a way that slows down the model. Furthermore, we propose using an additional layer of smoothing in portfolio construction post  Autoencoder portfolio weight retrieval. This could also be in the form of more sophisticated time series prediction techniques like LSTMs or Transformers trained in a suitable way.

A deeper analysis and economic interpretation of the resulting portfolios could offer insightful comparisons with PCA-derived or fundamental factors, particularly in understanding the impact of non-linearities.

We also aim to refine the architecture to dynamically learn the optimal number of latent factors to instead of setting the hyperparameter upfront. PCA research shows that the optimal number of factors also changes depending on the economic regime and we aim to include that information in our StatArb pipeline.

Finally, it is important to acknowledge that successful StatArb strategies often operate on intraday data due to the brief mean reversion times of stock residuals. This contrasts with the focus on interday data in much of the academic literature, pointing towards a significant area for further exploration in both practical and academic contexts.

\bibliography{statarb}
\bibliographystyle{icml2024}

\newpage

\appendix
\onecolumn
\section{Benchmark Results}
The following tables \ref{tab-ou-comp} and \ref{tab-ou-ffn-app} show the remaining benchmark and Autoencoder results that were referred to in the main text. Table \ref{tab-ou-comp} gives the full comparison of asset pricing models against the simple Autoencoder variants we derived using the OU signal extraction methodology and table \ref{tab-ou-ffn-app} compares the different models using the OU+FFN signal extraction methodology.

\begin{table}[bh]
\caption{Performance comparison of the benchmark asset pricing models against our simple Autoencoder architectures using Ornstein-Uhlenbeck time series extraction.}
\label{tab-ou-comp}
\vskip 0.15in
\begin{center}
\begin{small}
\begin{sc}
\scalebox{0.9}{
\begin{tabular}{l|c|ccc|ccc|r}
\toprule
Model & Variant & SR & $\mu$ & $\sigma$  \\
\midrule
\midrule 
\multirow{4}{*}{FF OU}  & CAPM & 0.13 & 1.31\% & 10.17\% \\ 
 & FF 3 & 0.3 & 2.07\% & 6.89\% \\ 
& FF 5 & 0.33 & 2.22\% & 6.78\% \\ 
& FF 5 Mom & 0.52 & 3.39\% & 6.56\% \\ 
\midrule
\midrule
\multirow{8}{*}{PCA OU}  & PCA 1 & 0.44 & 2.92\% & 6.64\% \\ 
& PCA 3 & 0.74 & 4.04\% & 5.48\% \\ 
& PCA 5 & 0.76 & 3.88\% & 5.07\% \\ 
& PCA 6 & 0.88 & 4.33\% & 4.93\% \\ 
& PCA 8 & 0.96 & 4.55\% & 4.74\% \\ 
& PCA 10 & 0.92 & 4.3\% & 4.66\% \\ 
& PCA 15 & 0.87 & 3.94\% & 4.53\% \\ 
& PCA 20 & 0.88 & 3.91\% & 4.43\% \\ 
\midrule
\midrule 
\multirow{10}{*}{AE OU 1} & Variant 0 & -0.24 & -3.17\% & 13.45\% \\ 
& Variant 1 & -0.29 & -3.91\% & 13.65\% \\ 
 & Variant 2 & -0.24 & -3.33\% & 13.84\% \\ 
 & Variant 3 & -0.23 & -3.27\% & 14.12\% \\ 
 & Variant 4 & -0.26 & -3.78\% & 14.39\% \\ 
 & Variant 5 & -0.18 & -1.65\% & 9.25\% \\ 
 & Variant 6 & -0.19 & -1.72\% & 9.24\% \\ 
 & Variant 7 & -0.24 & -2.42\% & 9.94\% \\ 
 & Variant 8 & -0.23 & -2.24\% & 9.78\% \\ 
 & Variant 9 & -0.18 & -1.66\% & 9.38\% \\
\midrule
\midrule 
\multirow{10}{*}{AE OU 2} & Variant 0 & 0.14 & 2.36\% & 16.75\% \\ 
 & Variant 1 & 0.34 & 5.76\% & 16.96\% \\ 
 & Variant 2 & 0.29 & 4.84\% & 16.77\% \\ 
 & Variant 3 & 0.31 & 5.17\% & 16.75\% \\ 
 & Variant 4 & 0.1 & 1.68\% & 17.25\% \\ 
 & Variant 5 & 0.23 & 3.83\% & 16.32\% \\ 
 & Variant 6 & 0.22 & 3.61\% & 16.1\% \\ 
 & Variant 7 & 0.25 & 3.93\% & 16.0\% \\ 
 & Variant 8 & 0.25 & 4.01\% & 16.03\% \\ 
& Variant 9 & 0.15 & 2.52\% & 16.45\% \\ 
\midrule
\midrule 
\multirow{10}{*}{AE OU 3} & Variant 0 & 0.17 & 2.93\% & 16.76\% \\ 
 & Variant 1 & 0.24 & 4.06\% & 17.2\% \\ 
 & Variant 2 & 0.27 & 4.44\% & 16.55\% \\ 
 & Variant 3 & 0.24 & 4.04\% & 16.56\% \\ 
 & Variant 4 & 0.14 & 2.35\% & 17.26\% \\ 
 & Variant 5 & 0.27 & 4.34\% & 16.36\% \\ 
 & Variant 6 & 0.42 & 6.93\% & 16.52\% \\ 
 & Variant 7 & 0.32 & 5.28\% & 16.31\% \\ 
 & Variant 8 & 0.32 & 5.18\% & 16.33\% \\ 
 & Variant 9 & 0.07 & 1.24\% & 16.98\% \\
\hline
\bottomrule
\end{tabular}}
\end{sc}
\end{small}
\end{center}
\vskip -0.1in
\end{table}

\begin{table}[ht]
\caption{Performance comparison of the benchmark asset pricing models against our simple Autoencoder architectures using the OU+FFN time series extraction.}
\label{tab-ou-ffn-app}
\vskip 0.15in
\begin{center}
\begin{small}
\begin{sc}
\scalebox{0.9}{
\begin{tabular}{l|c|ccc|ccc|r}
\toprule
Model & Variant & SR & $\mu$ & $\sigma$ \\
\midrule
\midrule 
\multirow{4}{*}{FF OU+FFN} & CAPM & 0.03 & 0.41\% & 15.1\% \\ 
 & FF 3 & -0.31 & -4.81\% & 15.67\% \\ 
& FF 5 & 0.11 & 1.65\% & 15.38\% \\ 
& FF 5 Mom & -0.09 & -1.38\% & 15.88\% \\
\midrule
\midrule
\multirow{8}{*}{PCA OU+FFN} & PCA 1 & -0.04 & -0.63\% & 17.27\% \\ 
 & PCA 3 & -0.1 & -1.65\% & 16.77\% \\ 
& PCA 5 & -0.18 & -2.69\% & 15.09\% \\ 
& PCA 6 & -0.19 & -3.48\% & 17.89\% \\ 
& PCA 8 & -0.22 & -3.37\% & 15.46\% \\ 
& PCA 10 & -0.17 & -2.59\% & 15.6\% \\ 
& PCA 15 & -0.31 & -4.88\% & 15.84\% \\ 
& PCA 20 & -0.08 & -1.41\% & 17.89\% \\ 
\midrule
\midrule 
\multirow{10}{*}{AE OU+FFN 2} & Variant 0 & 0.08 & 1.64\% & 19.56\% \\ 
 & Variant 1 & 0.06 & 1.14\% & 20.3\% \\ 
 & Variant 2 & -0.09 & -1.75\% & 19.21\% \\ 
 & Variant 3 & 0.06 & 1.28\% & 20.32\% \\ 
 & Variant 4 & 0.08 & 1.54\% & 20.36\% \\ 
 & Variant 5 & 0.12 & 2.15\% & 18.43\% \\ 
 & Variant 6 & -0.3 & -5.3\% & 17.83\% \\ 
 & Variant 7 & 0.04 & 0.72\% & 18.07\% \\ 
 & Variant 8 & 0.01 & 0.13\% & 18.34\% \\ 
 & Variant 9 & -0.29 & -5.37\% & 18.33\% \\ 
\hline
\bottomrule
\end{tabular}}
\end{sc}
\end{small}
\end{center}
\vskip -0.1in
\end{table}

\end{document}